\documentclass[aps,prl,twocolumn,amsmath,showpacs,superscriptaddress]{revtex4-1}
\usepackage{graphicx,epsfig,longtable}
\usepackage{epstopdf}
\usepackage{dcolumn}
\usepackage{bm}

\def\ga{\,\,\raise0.14em\hbox{$>$}\kern-0.76em\lower0.28em\hbox
{$\sim$}\,\,}
\def\la{\,\,\raise0.14em\hbox{$<$}\kern-0.76em\lower0.28em\hbox
{$\sim$}\,\,}

\def\Msun{$M_{\odot}$}

\begin{document}
\title{New fission fragment distributions and r-process origin of the rare-earth elements}

\author{S.~Goriely}
\affiliation{Institut d'Astronomie et d'Astrophysique, CP-226, Universit\'e Libre de Bruxelles, 
1050 Brussels, Belgium}
\author{J.-L. Sida}
\author{J.-F. Lema\^\i tre}
\author{S. Panebianco}
\affiliation{C.E.A. Saclay, Irfu/Service de Physique Nucl\'eaire, 91191 Gif-sur-Yvette, France}
\author{N. Dubray}
\author{S. Hilaire}
\affiliation{CEA, DAM, DIF, F-91297 Arpajon, France}
\author{A. Bauswein}
\affiliation{Department of Physics, Aristotle University of Thessaloniki, 54124 Thessaloniki, Greece}
\affiliation{Max-Planck-Institut f\"ur Astrophysik, Postfach 1317, 85741 Garching, Germany}
\author{H.-T. Janka}
\affiliation{Max-Planck-Institut f\"ur Astrophysik, Postfach 1317, 85741 Garching, Germany}

\date{\today}


\begin{abstract}
Neutron star (NS) merger ejecta offer a viable site for the
production of heavy r-process elements with nuclear mass numbers
$A\ga 140$. The crucial role of fission recycling is responsible
for the robustness of this site against many astrophysical uncertainties,
but calculations sensitively depend on nuclear physics.
In particular the fission fragment yields determine the 
creation of $110\la A\la 170$ nuclei. Here we apply a new scission-point
model, called SPY, to derive the fission fragment distribution (FFD)  of all relevant neutron-rich, 
fissioning nuclei. The model predicts a doubly asymmetric FFD
in the abundant $A\simeq 278$ mass region that is responsible for the final 
recycling of the fissioning material. Using ejecta conditions based on
relativistic NS merger calculations we show that this specific FFD leads
to a production of the $A\simeq 165$ rare-earth peak that is nicely
compatible with the abundance patterns in the Sun and metal-poor stars.
This new finding further strengthens the case of NS mergers as possible
dominant origin of r-nuclei with $A\ga 140$.
\end{abstract}



\maketitle



\maketitle

{\em Introduction.}---The
rapid neutron-capture process (r-process) of stellar nucleosynthesis
explains the production of the stable  (and some long-lived radioactive) neutron-rich nuclides 
heavier than iron that are observed in stars of various metallicities and in the solar
system (see review of \cite{ar07}).
While r-process theory has made progress in understanding possible mechanisms that could be at the origin of the solar-system composition, the cosmic site(s) of the r-process has (have) not been identified yet and the astrophysical sources and specific conditions in which  the
r-process takes place are still among the most longstanding mysteries of nuclear astrophysics.

Progress in modelling core-collapse supernovae (SNe) and $\gamma$-ray bursts has raised a lot of excitement about the so-called
neutrino-driven wind environment \cite{ar07,wanajo11,janka12}. While the light r-elements up to the second 
abundance peak ($A\sim 130$) might be produced in such outflows of nascent neutron stars (NSs)\cite{wanajo11,arcones11},
the extreme conditions required for stronger r-processing have so far not been
obtained in the most sophisticated SN models \cite{janka12}. An alternative to the r-process  in high-temperature SN environments is the decompression of cold neutronized matter from violent collisions of binary NSs or NSs with companion black holes.  While  such a connection was suggested decades ago~\cite{lattimer74,lattimer76,eichler89} and decompressed NS matter was found to be favorable for strong r-processing~\cite{meyer89},
only more recent and increasingly sophisticated hydrodynamic simulations could
determine the ejecta mass to be $\sim$$10^{-3}$--$10^{-2}$\,\Msun~\cite{ruffert96,ruffert97,rosswog99,frei99,roberts11,rosswog13a,rosswog13b,hotokezaka13,goriely11,bauswein13}.  
This mass, combined with the predicted astrophysical event rate ($\sim$$10^{-5}$\,yr$^{-1}$ in the Milky Way~\cite{dominik12,kalogera04}) can account for the majority of r-material in our Galaxy \cite{ruffert97,frei99,goriely11,bauswein13,qian00,metzger10}.
Nearly all of the ejecta are converted to r-process nuclei, whose radioactive decay heating leads to potentially observable electromagnetic radiation in the optical and infrared bands~\cite{metzger10,li98} with 100--1000 times fainter peak brightnesses than those of typical SNe and durations of only days~\cite{roberts11,goriely11,bauswein13,barnes13,tanaka13,grossman13}. These ``macronovae'' \cite{kulkarni05} or ``kilonovae'' \cite{metzger10} are intensely searched for (with a recent, possible first success~\cite{berger13,tanvir13}) and their unambiguous discovery would constitute the first detection of r-material in situ.

In this specific r-process scenario, the number of free neutrons per seed nucleus reach a few hundreds. With such a neutron richness, fission plays a fundamental role by recycling the matter during the neutron irradiation and by shaping the final r-abundance distribution in the $110 \la A \la 170$ mass region at the end of the neutron irradiation. The final composition of the ejecta is then rather insensitive to details of the initial abundances and the astrophysical conditions, in particular the mass ratio of the two NSs, the quantity of matter ejected, and the equation of state (EOS)~\cite{goriely11,bauswein13,korobkin12}. This robustness, which is compatible with the uniform, solar-like abundance pattern of the rare-earth elements observed in metal-poor stars~\cite{sneden08}, might point to the creation of these elements by fission recycling in NS merger (NSM) ejecta. 

However, the estimated abundance distribution remains sensitive to the adopted nuclear models.
The ejecta are composed almost exclusively of $A>140$  nuclei, and in particular the $A\simeq 195$ third r-process peak appears in proportions similar to those observed in the solar system, deviations resulting essentially from the  still difficult task to predict neutron capture and $\beta$-decay rates for exotic neutron-rich nuclei. The situation for the lighter $110 \la A \la 170$ species has been rather unclear up to now and extremely dependent on fission properties, including in particular the fission fragment distribution (FFD).  In the present paper, we apply a new state-of-the-art  scission-point model, called SPY, to the determination of the FFD of all neutron-rich fissioning nuclei of relevance during the r-process nucleosynthesis and analyze its impact on the r-process abundance distribution.

{\em NS merger simulations and the r-process.}---Our
NSM simulations were performed with a general relativistic Smoothed 
Particle Hydrodynamics scheme \citep{oech07,bauswein10,bauswein13} representing the fluid by 
a set of particles with constant rest mass, whose properties were 
evolved according to Lagrangian hydrodynamics, conserving the electron fraction of fluid elements. The Einstein field equations were
solved assuming a conformally flat spatial metric.  The r-abundance distributions resulting from binary simulations with different mass ratios or different EOSs are virtually identical \cite{bauswein13}. 
For this reason, in the present analysis only symmetric 1.35\Msun--1.35\Msun\ systems with the DD2 EOS \cite{typel10,hempel10}, including thermal effects and a resolution of $\sim$550,000 particles, are considered. The mass ejected by the NSM is $\sim$$3\times 10^{-3}$\,\Msun. In \cite{bauswein10,bauswein13} more details are given on gross properties of the ejecta, the influence of the EOS and the postprocessing for the nucleosynthesis calculations.  
Note that the 1.35\Msun--1.35\Msun\ case is of particular interest since, according to population synthesis studies and pulsar observations, it represents the most abundant systems \citep{bel08}. 

Our nuclear network calculations were performed as in~\cite{goriely11,goriely11b}, where 
the reaction network, temperature postprocessing, inclusion of pressure feedback by nuclear
heating, and the density extrapolation beyond the end of the hydrodynamical simulations are described.
The reaction network includes all 5000 species from protons up to Z = 110 that lie
between the valley of $\beta$-stability and the neutron-drip line. All fusion reactions on
light elements as well as radiative neutron captures, photodisintegrations, $\alpha$- and
$\beta$-decays, and fission processes, are included. The corresponding rates are based on
experimental data whenever available or on theoretical predictions otherwise, as obtained from
the BRUSLIB nuclear astrophysics library \citep{xu13}. In particular, the reaction rates
are estimated with the TALYS code~\citep{ko05,go08} on the basis of the Skyrme
Hartree-Fock-Bogolyubov (HFB) nuclear mass model, HFB-21~\citep{gcp10}, and the
$\beta$-decays with the Gross Theory 2 (GT2)~\citep{tachibana90}, employing the same HFB-21
$Q$-values.

\begin{figure}
\includegraphics[scale=0.28]{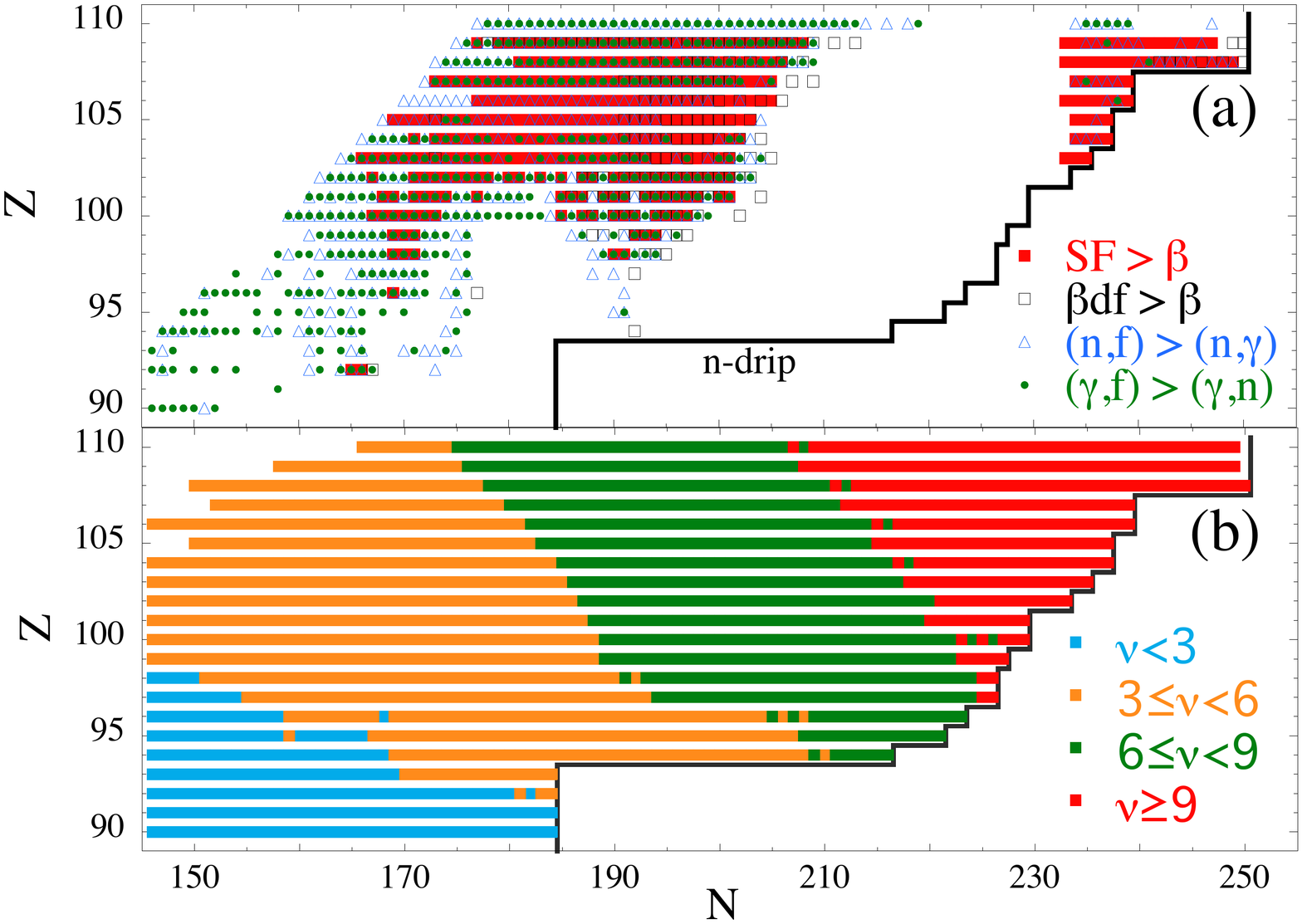}
\caption{ (a) Dominant fission regions in the $(N,Z)$ plane. 
Nuclei with spontaneous fission being faster than $\beta$-decays are shown 
by full squares, those with $\beta$-delayed fission faster than 
$\beta$-decays by open squares, those with neutron-induced fission 
faster than radiative neutron capture at $T=10^9\,$K by open triangles,
and those for which photo-fission at $T=10^9\,$K is faster than photo-neutron 
emission by closed circles. For $Z=110$, $\beta$-decay processes are not 
calculated.
(b) SPY predictions of the average number of emitted neutrons for each fissioning 
nucleus in the $(N,Z)$ plane.}
\label{fig01}
\end{figure}

The neutron-induced, photo-induced, $\beta$-delayed and spontaneous fission rates are estimated on the basis of the HFB-14 fission paths~\citep{go07}. The neutron- and photo-induced fission rates were calculated with the TALYS code for all nuclei with $90 \le Z \le 110$~\citep{go09}. Similarly, the $\beta$-delayed and spontaneous fission rates are estimated with the same TALYS fission barrier penetration calculation. The $\beta$-delayed fission rate takes into account the full competition between the fission, neutron and photon channels, weighted by the population probability given by the $\beta$-decay strength function~\citep{kodoma75}. 
The main fission regions by one of the four fission processes are illustrated in Fig.~\ref{fig01}a.

{\em SPY fission fragment distribution.}---To
study precisely the impact of the nascent fragment nuclear structure on the mass distribution, a renewed statistical scission-point model, called SPY, was developed~\cite{panebianco12}.
It consists of a parameter-free approach based on up-to-date microscopic ingredients extracted with a mean-field description using the effective nucleon-nucleon Gogny interaction~\cite{hilaire07}. This renewed version of the Wilkins fission model~\cite{wilkins76} estimates first the absolute energy available for all possible fragmentations at the scission point for a given fissioning nucleus~\cite{panebianco12}. The main ingredient in these calculations is the individual potential energy of each fission fragment as a function of its axial deformation, as compiled in the AMEDEE database~\cite{hilaire07} for more than 8000 nuclei. 
 Once the available energies are calculated for each fragmentation, a microcanonical description including nuclear Fermi gas state densities is used to determine the main fission fragment observables, more particularly mass and charge yields, kinetic energy and excitation energy of the fragments~\cite{lemaitre13}. The number of evaporated neutrons is deduced from the mean excitation energy of each fragment. The scission-point models~\cite{wilkins76} have shown their ability to reproduce the general trends of the fission yields for actinides, and the SPY model has proven its capability to describe them up to exotic nuclei in the study of the mercury isotopes~\cite{panebianco12}.

\begin{figure}
\includegraphics[width=9cm,height=6cm]{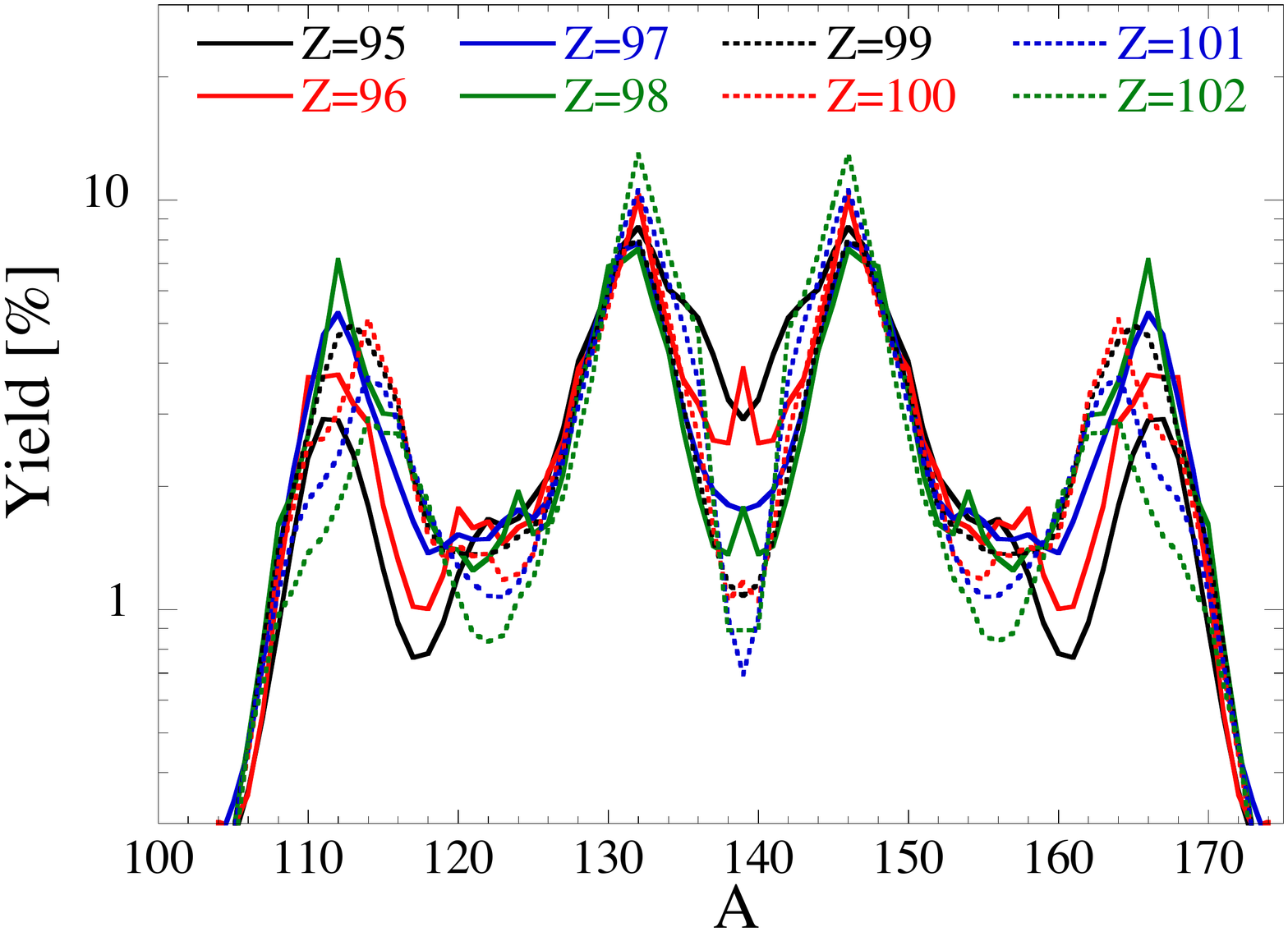}
\vskip -0.5 cm
\caption{ FFDs from the SPY model for eight $A=278$ isobars. } 
\label{fig02}
\end{figure}

\begin{figure}
\includegraphics[scale=0.3]{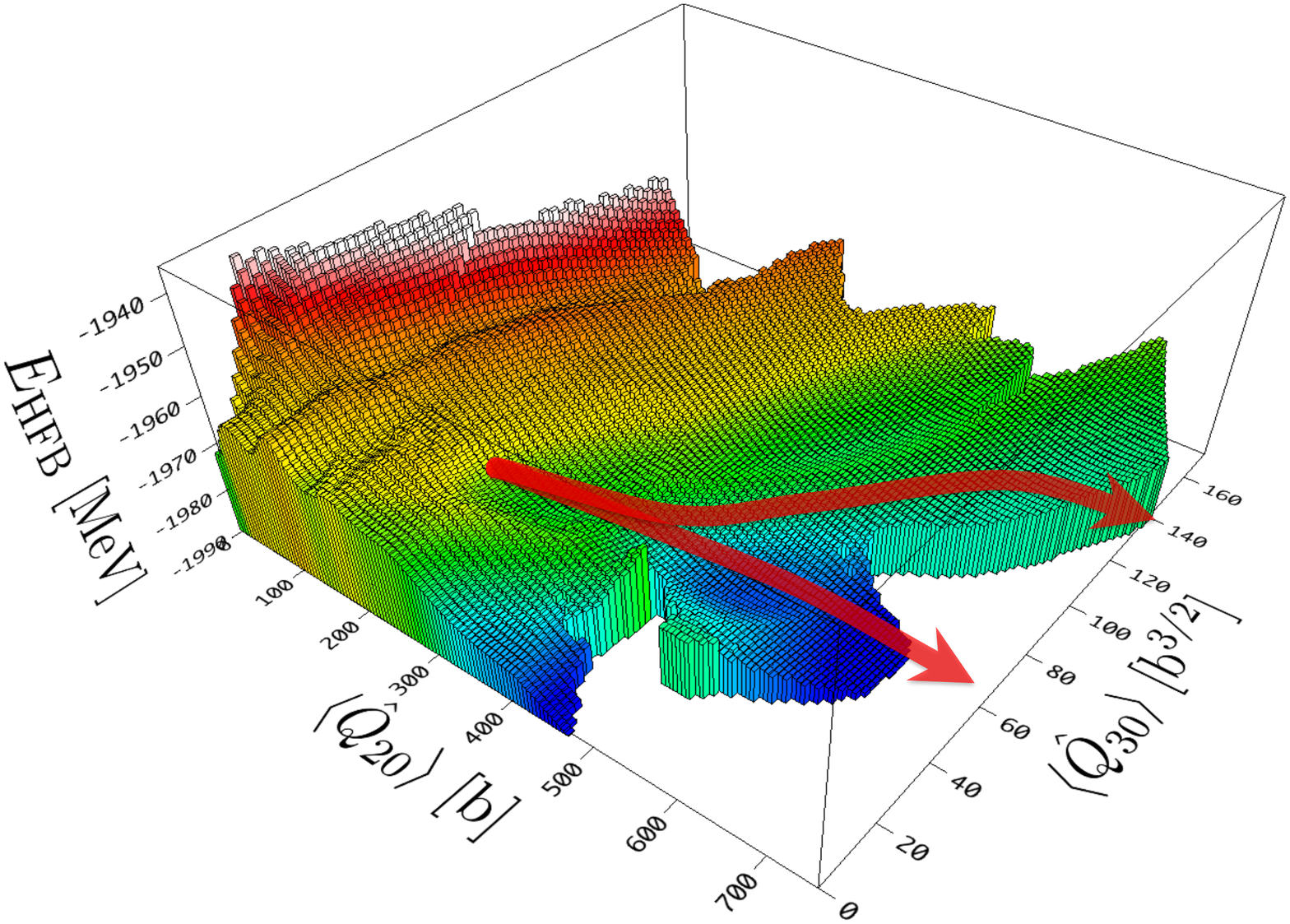}
\vskip -0.5 cm
\caption{$^{278}$Cf potential energy surface as a function of the quadrupole $\langle \hat Q_{20}\rangle $ and octupole $\langle \hat Q_{30}\rangle $ deformations. 
Both asymmetric fission valleys are depicted by the red arrows.} 
\label{fig03}
\end{figure}

SPY has now been applied to all the neutron-rich nuclei of relevance for r-process nucleosynthesis. 
It is found that the $A\simeq 278$ fissioning nuclei, which are main progenitors of the $110 \la A \la 170$ nuclei in the decompression of NS matter, present an unexpected doubly asymmetric fission mode with a characteristic four-hump pattern, as illustrated in Fig.~\ref{fig02}. Such fragment distributions have never been observed experimentally and can be traced back to the predicted potential energies at large deformations of the neutron-rich fragments favored by the $A\simeq 278$ fission. The two asymmetric fission modes can also be seen on the potential energy surface (Fig.~\ref{fig03}) obtained from a detailed microscopic calculation \cite{dubray08} for $^{278}$Cf in the deformation subspace (elongation $\langle
\hat{Q}_{20}\rangle$, asymmetry $\langle \hat{Q}_{30}\rangle$). This calculation uses a state-of-the-art
mean-field model with the Gogny interaction. The two fission valleys indicated by arrows in Fig.~\ref{fig03} lead  to asymmetries similar to the
distributions presented in Fig.~\ref{fig02} obtained with SPY. The symmetric valley, corresponding to a nil octupole moment, is disfavored by a smaller barrier transmission probability linked to the presence of a
barrier, hidden in this subspace by a discontinuity \cite{dubray12}. 

Finally, we show in Fig.~\ref{fig01}(b) the SPY prediction of the average number of evaporated neutrons for each spontaneously fissioning nucleus. This average number is seen to reach values of about four for the $A\simeq 278$ isobars and maximum values of $\sim$14 for the heaviest $Z\simeq110$ nuclei lying at the neutron drip line.

\begin{figure}
\includegraphics[scale=0.30]{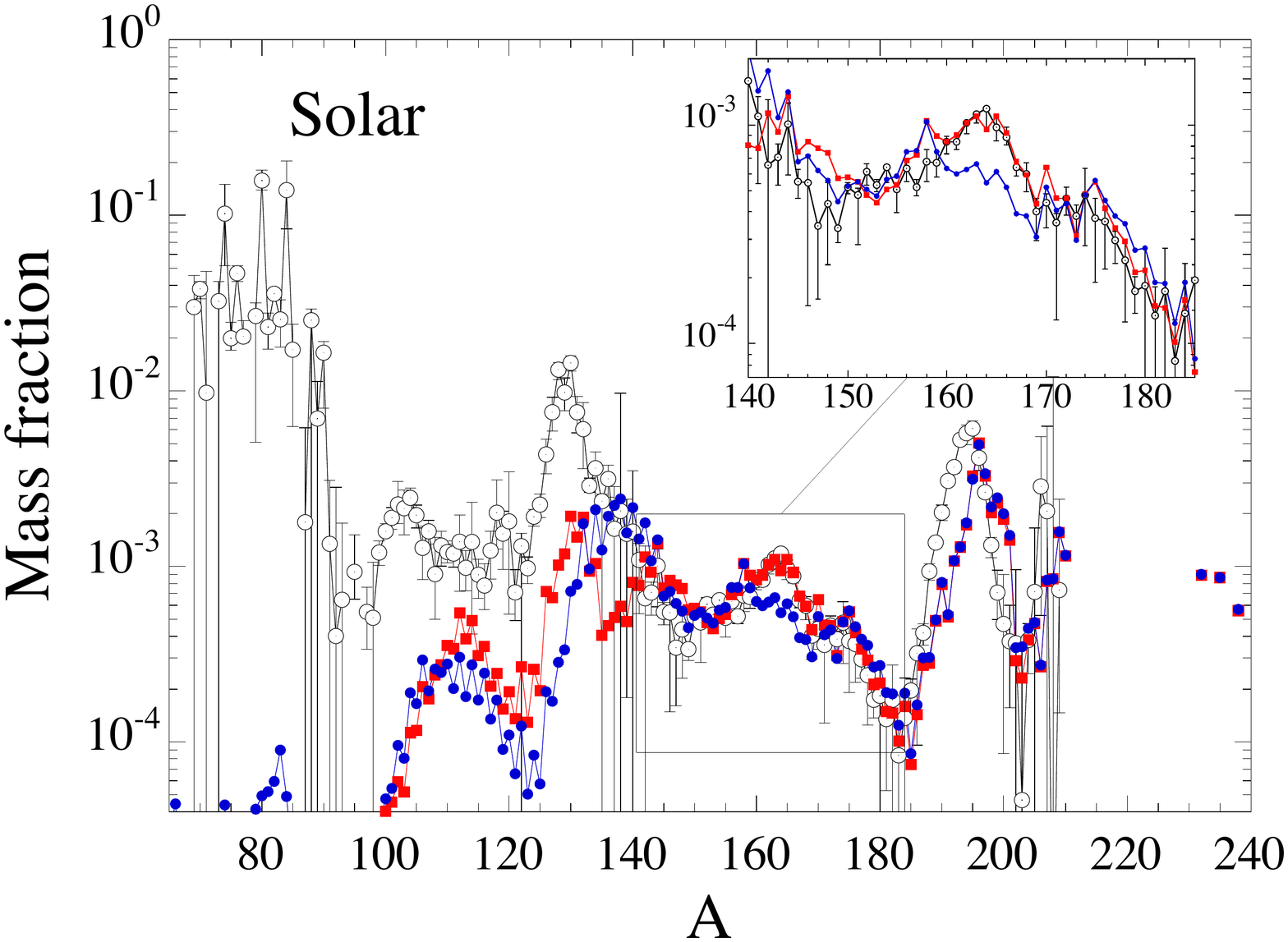}
\vskip -0.5 cm
\caption{Final abundance distribution vs.\ atomic mass for ejecta from 
1.35--1.35~M$_{\odot}$ NS mergers. The red squares are for the newly derived SPY
predictions of the FFDs and the blue circles for essentially symmetric 
distributions based on the 2013 GEF model~\cite{gef10}. The abundances are
compared with the solar ones \cite{go99} (dotted circles). The insert 
zooms on the rare-earth elements.}
\label{fig04}
\end{figure}

{\em Nucleosynthesis calculations.}---Due
to the specific initial conditions of high neutron densities (typically  $N_n\simeq 10^{33-35} {\rm cm^{-3}}$ at the drip density), the nuclear flow  during most of the neutron irradiation will follow the neutron-drip line and produce in milliseconds the heaviest drip-line nuclei. However, for drip-line nuclei with $Z\ge 103$, neutron-induced and spontaneous fission become efficient (Fig.~\ref{fig01}a) prohibiting the formation of super-heavy nuclei and recycling the heavy material into lighter fragments, which restart capturing the free neutrons. Fission recycling can take place up to three times before the neutrons are exhausted, depending on the expansion timescales. 
When the neutron density drops below some $10^{20}$~cm$^{-3}$, the timescale of neutron capture becomes longer than a few seconds, and the nuclear flow is dominated by $\beta$-decays back to the stability line (as well as fission and $\alpha$-decay for the heaviest species).  
The final abundance distribution of the $3\times 10^{-3}$\Msun\ of ejecta during the NSM is compared with the solar system composition in Fig.~\ref{fig04}. The similarity between the solar abundance pattern and the prediction in the $140 \la A \la 180$ region is remarkable and strongly suggests that this pattern constitutes the standard signature of r-processing under fission conditions.

The $110 \la A \la 170$ nuclei originate exclusively from the spontaneous and $\beta$-delayed fission recycling that takes place in the $A\simeq 278$ region at the time all neutrons have been captured and the $\beta$-decays dominate the nuclear flow. The $A\simeq 278$ isobars correspond to the dominant abundance peak in the actinide region during the irradiation phase due to the turn-off point at the $N=184$ drip-line shell closure and the bottleneck created by $\beta$-decays along the nuclear flow. The nuclei that $\beta$-decay  along the $A=278$ isobar fission asymmetrically according to the SPY FFD model,  as illustrated in Fig.~\ref{fig02}, leading to a similar quadruple hump pattern visible in Fig.~\ref{fig04} (red squares).  The asymmetric $A\simeq 165$ peak in the FFD (Fig.~\ref{fig02}) can consequently explain the origin of the rare-earth peak by the r-process, in contrast to more phenomenological FFD models~\citep{kodoma75}, which predict symmetric mass yields for the $A\simeq 278$ fissioning nuclei and hence an underproduction of the $A\simeq 165$ rare-earth nuclei (cf.\ Fig.~4 in~\cite{goriely11}). An essentially symmetric FFD is also predicted by the 2013 version of the semi-empirical GEF model~\cite{gef10}, also leading to an underproduction of rare-earth elements, as shown in Fig.~\ref{fig04} and also discussed in Ref.~\cite{goriely13}. Our NSM scenario thus offers a consistent explanation of the creation of the rare-earth elements connected to r-processing, different from alternative suggestions for production sites of these elements, e.g.\ at freeze-out conditions in high-entropy r-process environments~\citep{mumpower12} with all the associated astrophysical problems~\cite{ar07,wanajo11,janka12}.

In addition, with the SPY FFDs the r-abundance distribution is rather robust for different sets of fission barriers. As explained above, the $110 \la A \la 170$ abundances originate essentially from the fission of the nuclei that $\beta$-decay along the $A\simeq 278$ isobars at the end of the neutron irradiation. The corresponding fissioning nuclei are all predicted by the SPY model to fission basically with the same doubly asymmetric distribution (Fig.~\ref{fig02}), leading to similar r-distributions, independent of the fissioning element along the isobar. 

\begin{figure}
\includegraphics[scale=0.28]{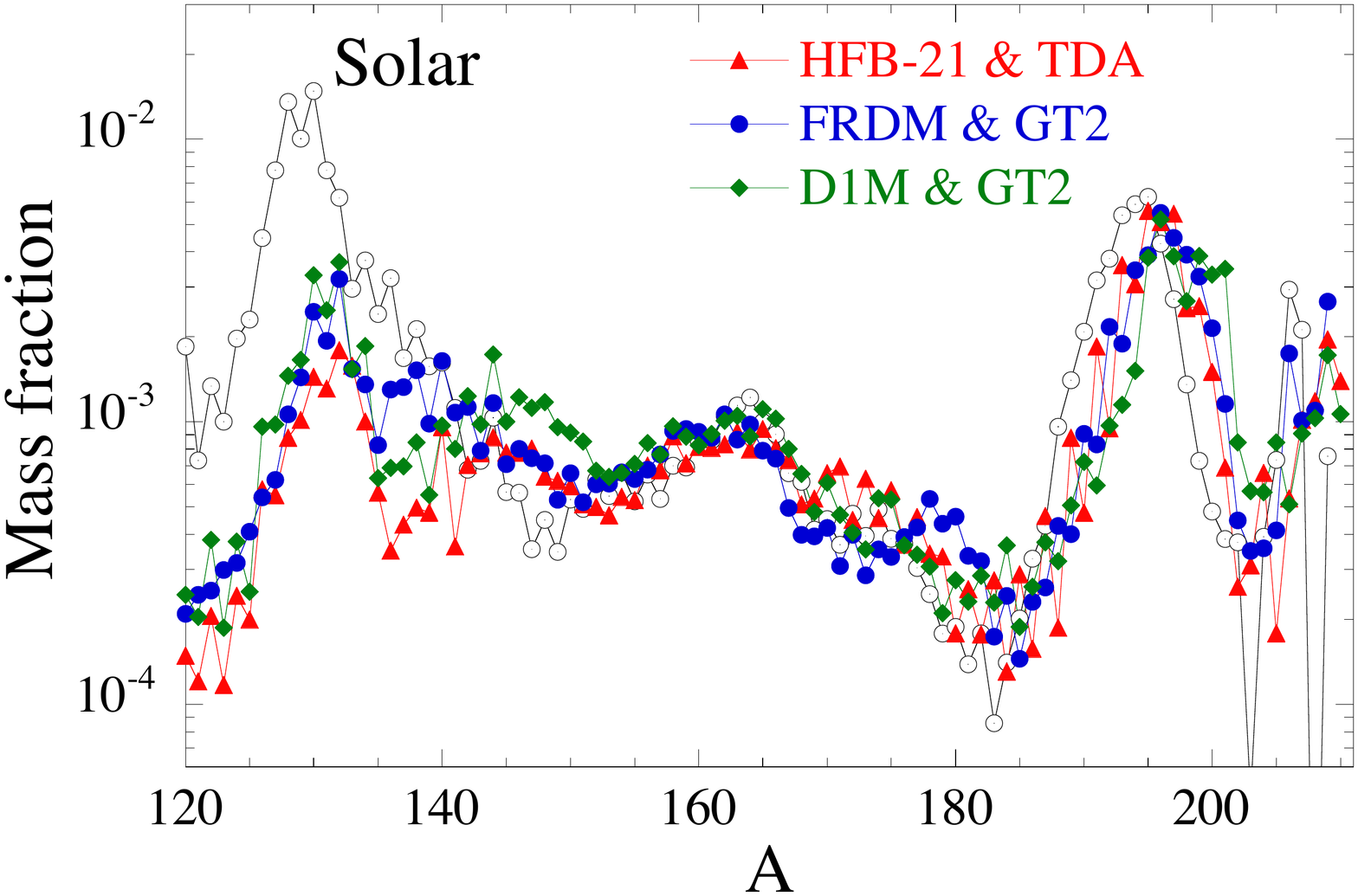}
\vskip -0.5 cm
\caption{Same as Fig.~\ref{fig04} but with abundance distributions obtained with
 three additional sets of nuclear rates, namely reaction rates obtained with  
the D1M~\cite{gor09} or FRDM~\cite{frdm} masses and $\beta$-decay rates 
from the GT2 or Tamm-Dancoff approximation (TDA)~\cite{klapdor84}.} 
\label{fig05}
\end{figure}

The emission of prompt neutrons also affects the r-abundance distribution. According to the SPY model, the fission of the most abundant nuclei around $A=278$ is accompanied with the emission of typically 4 neutrons (Fig.~\ref{fig01}b). These neutrons are mainly re-captured by the abundant nuclei forming the $N=126$ peak. For this reason, not only the abundance distribution for $A \la 160$ is slightly shifted to lower masses, but the abundant $A=196$ peak  is shifted to higher masses  by a few units. The impact, however, remains small due to the small average number of emitted neutrons. This even improves the agreement with the solar distribution for $A\simeq 145$ and $A\simeq 172$ nuclei but distorts slightly the $A=195$ peak. However, 
the global abundance pattern for $A>140$, in particular the $A=195$ peak, can also be affected by the still uncertain neutron-capture and $\beta$-decay rates. Nevertheless, the production of the rare-earth peak remains qualitatively rather robust (Fig.~\ref{fig05}), at least for the three additional sets of nuclear models tested here. 

{\em Conclusions.}---The decompression of NS matter remains a promising site for the r-process. This site is extremely robust with respect to many astrophysical uncertainties. We demonstrated here that the newly derived FFD based on the SPY model can consistently explain the abundance pattern in the rare-earth peak within this r-process scenario, in contrast to results with more phenomenological models predicting symmetric mass yields for the fissioning $A\simeq 278$ nuclei. Our new finding provides an even stronger hint to NSMs as possibly dominant site for the origin of $A>140$ r-nuclei in the Universe. In particular the robustness of the ejecta conditions and associated fission recycling as well as the good quantitative agreement of the theoretical and solar abundances are fully compatible with the amazing uniformity of the rare-earth abundance patterns observed in many metal-poor stars~\cite{sneden08}.

The unexpected doubly asymmetric FFD predicted by SPY also opens new perspectives in theoretical and experimental nuclear physics concerning specific fission modes related to the nuclear structure properties of exotic nuclei. Dynamical mean field calculations~\cite{goutte05} should quantitatively confirm the fission yields predicted by SPY, and future experiments producing fission fragments similar to those predicted by the doubly asymmetric fission mode could reveal the nuclear properties of the corresponding fission fragments. 

\begin{acknowledgements}
S.G. acknowledges financial support of F.N.R.S. and ``Actions de recherche concert\'ees (ARC)'' from the ``Communaut\'e fran\c caise de Belgique''. A.B. and H.-T.J. acknowledge support by Deutsche Forschungsgemeinschaft through the Transregional Collaborative Research Center SFB/TR~7 ``Gravitational Wave Astronomy'' and the Cluster of Excellence EXC~153 ``Origin and Structure of the Universe''. A.B. is Marie Curie
Intra-European Fellow within the 7th European Community Framework Programme (IEF 331873)
\end{acknowledgements}

\end{document}